\documentclass[aps,prd,preprint,tightenlines,superscriptaddress,nofootinbib]{revtex4}
\usepackage{epsfig}
\usepackage{graphicx}
\usepackage{psfrag}
\usepackage{amsmath,amssymb}
\usepackage{colordvi}
\usepackage{amsfonts}
\usepackage{enumerate}
\usepackage{slashed}
\usepackage{color}
\usepackage{xcolor}

\begin{document}

\title{Scintillation Efficiency for Low-Energy Nuclear Recoils \\ in Liquid-Xenon Dark Matter Detectors}

\author{Wei Mu}
\affiliation{INPAC, Department of Physics, Shanghai Jiao Tong University, Shanghai, 200240, P. R. China}
\author{Xiaonu Xiong}
\affiliation{Center for High-Energy Physics, Peking University, Beijing, 100080, P. R. China}
\author{Xiangdong Ji}
\affiliation{INPAC, Department of Physics, Shanghai Jiao Tong University, Shanghai, 200240, P. R. China}
\affiliation{Center for High-Energy Physics, Peking University, Beijing, 100080, P. R. China}
\affiliation{Maryland Center for Fundamental Physics, Department of Physics, University of Maryland, College Park, Maryland 20742, USA}
\date{\today}
\vspace{0.5in}

\begin{abstract}
We perform a theoretical study of the scintillation efficiency in the low-energy region crucial for liquid-xenon dark-matter detectors. We develop a computer program to simulate the cascading process of the recoiling xenon nucleus in liquid xenon and calculate the nuclear quenching effect due to atomic collisions. We use the electronic stopping power extrapolated from the experimental data to the low-energy region, and take into account the effects of electrons' escaping from the electron-ion pair recombination using the generalized Thomas-Imel model fitted to scintillation data. Our result agrees well with the experiments from neutron scattering and vanishes rapidly as the recoiling energy drops below 3~keV.
\end{abstract}

\maketitle

There are myriad astrophysical evidences for the existence of the so-called dark matter~(DM) in the universe~\cite{Bertone:2004pz}. The most attractive candidate is the weakly interacting massive particles~(WIMPs): electrically-neutral stable particles with masses ranging from GeV to TeV with the standard-model-like weak interactions. WIMPs can be detected directly by observing the atomic recoils after their elastic scattering on nuclei. Many direct detection experiments have been proposed and run in the last two decades~\cite{Akimov:2011za}, among which liquid xenon~(LXe) detectors have shown particular promise. The recent XENON100 experiment has yielded the best detection limits in almost all regions of the possible WIMP masses, more sensitive than most other experiments using alternative detection media~\cite{Aprile:2012nq}. A list of xenon experiments, XMASS, LUX, PandaX, and XENON1t will soon join the direct search effort.

A crucial piece of information for LXe experiments is the so-called {\it scintillation efficiency}, required for the energy calibration. The detector independent value {\it relative scintillation efficiency} ${\cal L}_{\rm eff}$ has been widely used to relate scintillation signal to nuclear recoil energy~\cite{Aprile:2012nq},
\begin{equation}
{\cal L}_{\rm eff} (E_{\rm nr})=\frac{S_{\rm ee}}{S_{\rm nr}} \frac{S1(E_{\rm nr})}{E_{\rm nr}} \frac{1}{L_y} \ ,
\end{equation}
where $S1(E_{\rm nr})$ is the scintillation signal from nuclear recoil with initial energy $E_{\rm nr}$ while $L_y$ is that per keV from 122~keV electron recoils~($^{57}$Co $\gamma$ calibration line as the standard candle). The coefficient $S_{\rm ee}/S_{\rm nr}$ corrects for the finite external electric field applied to the detector so that ${\cal L}_{\rm eff}$ is drift-field independent.

A series of experiments~\cite{Aprile:2009pc,Manzur:2010,Plante:2011} have been done to measure ${\cal L}_{\rm eff}$ using the mono-energy neutron sources. However, for very low energy nuclear recoils, the measurements are particularly difficult. ${\cal L}_{\rm eff}$ has to be extrapolated to lower energy from the lowest measured energy 3~keV~\cite{Plante:2011} as a constant. Because of a kinematic cutoff, Collar and McKinsey~\cite{Collar:2010gg} pointed out the LXe scintillation response should drop steadily at low-energy where most of the recoiling events will be (2~keV or less) if the mass of WIMPs is around 10 GeV~\cite{Bernabei:2008yi,Aalseth:2010vx,Agnese:2013rvf}. On the theoretical side, a common practice is to firstly apply the Lindhard factor~(nuclear quenching factor $q_{\rm nc}(E_{\rm nr})=\eta (E_{\rm nr})/E_{\rm nr}$)~\cite{Lindhard:1963} to estimate the fraction of the energy given to electrons $\eta (E_{\rm nr})$ and then to calculate ${\cal L}_{\rm eff}$~\cite{Hitachi:2005,Mei:2007jn,Bezrukov:2011}. Simulation programs, such as TRIM~\cite{Ziegler:2010}, can also be used to derive $q_{\rm nc}$ in a medium. However, likewise they have not been calibrated for low-energy recoils. In particular, the threshold kinetic energy for atoms participating further cascade is used independent of the low-energy behavior of the electronic stopping power.

In this work, we aim to develop a state-of-art theoretical approach to simulate the recoiling-xenon's slowing down process in LXe media and get a realistic $q_{\rm nc}$ in the low-energy region. For the input electronic stopping power~(ESP), we use an extrapolated result from an educated fit to the experimental data in 40$\sim$100~keV~\cite{Fukuda:1981}. For the scintillation quenching effects, we treat electrons' escaping from electron-ion pair recombination by extending the Thomas-Imel model~\cite{Thomas:1987pa} to zero-external field, and fit to the experimental linear-energy-transfer~(LET, = ESP divided by density) dependence in LXe to find the free parameters. Combining all these, we get the (relative) scintillation efficiency at keV-recoiling energy region. The result compares favorably with the existing data from mono-energy neutron scattering and indicates a rapidly vanishing scintillation from the nuclear recoils below 3~keV.

Unlike electron recoils whose kinetic energies are all dissipated into electronic energy of the media, the energy dissipation process of nuclear recoils is more complicate. When a xenon nucleus is scattered by a DM particle and recoils inside the LXe media, atomic motions, electronic excitations and ionizations are generated, and the whole process is in principle a many-body quantum mechanics problem. To find the equivalent electronic energy dissipation to electron recoils, it is necessary to make reasonable simplifications to render a theoretical treatment feasible. Similar to previous works, we treat the nucleus-nucleus interactions classically as binary elastic collisions because their de Broglie wavelength (less than 0.0025 \AA) in the concerned energy region (0.5$\sim$25~keV) are much smaller than the atomic scale. In the meanwhile, the nucleus-electron interaction is treated as a small ``friction'' on the projectile along the nucleus trajectory. This is because the atomic scattering generates large momentum transfer, hence influence the atom trajectory but dissipate no kinetic energy, whereas the atom-electron interaction generates little momentum transfer but appreciable kinetic energy dissipation which can be considered as a small perturbation to atomic motion. The classical trajectories of nucleus are determined by the screened Coulomb potential $U(r)$,
\begin{equation}
U(r)=\frac{Z^2 e^2}{r} \Phi\left(\frac{r}{a}\right) \ ,
\end{equation}
where $\Phi(\tfrac{r}{a})$ is the Hartree-Fock screening function and $a$ is the screening radius for xenon, both of which can be found in Ref.~\cite{Ziegler:2010}. The binary collisions along the trajectories are determined by the spatial distribution of xenon atoms in liquid which has been measured in Ref.~\cite{Wamba:2007}. The electronic energy dissipation in a nuclear binary collision is determined by,
\begin{equation}
\label{eq:LFactor}
\eta (E_{\rm nr}^i)=\int_0^{\lambda(E_{\rm nr}^i)}S_e (E) dx \ ,
\end{equation}
where $\lambda(E_{\rm nr}^i)$ is the free flight path length of a recoiling atom with energy $E_{\rm nr}^i$ and $S_e (E)$ is ESP, the energy loss to electrons per unit path, $(dE/dx)_{\rm el}$. Transferring energy to other nuclei and dissipating energy to electrons, the moving particle slows down and the trajectory will be terminated when its kinetic energy approaches the thermal energy of atoms in LXe (0.02~eV). Notes, if the kinetic energy of the initial projectile xenon is lower than the gap energy in LXe (9.3~eV), we will not initialize the collision cascade because no electronic excitations or ionizations will be produced; the kinematic cutoff discussed in Ref.~\cite{Collar:2010gg} and in TIRM is reflected through a fast fall-off of the ESP at low energy in this work.

\begin{figure}[hbt]
\begin{center}
\includegraphics[width=8cm]{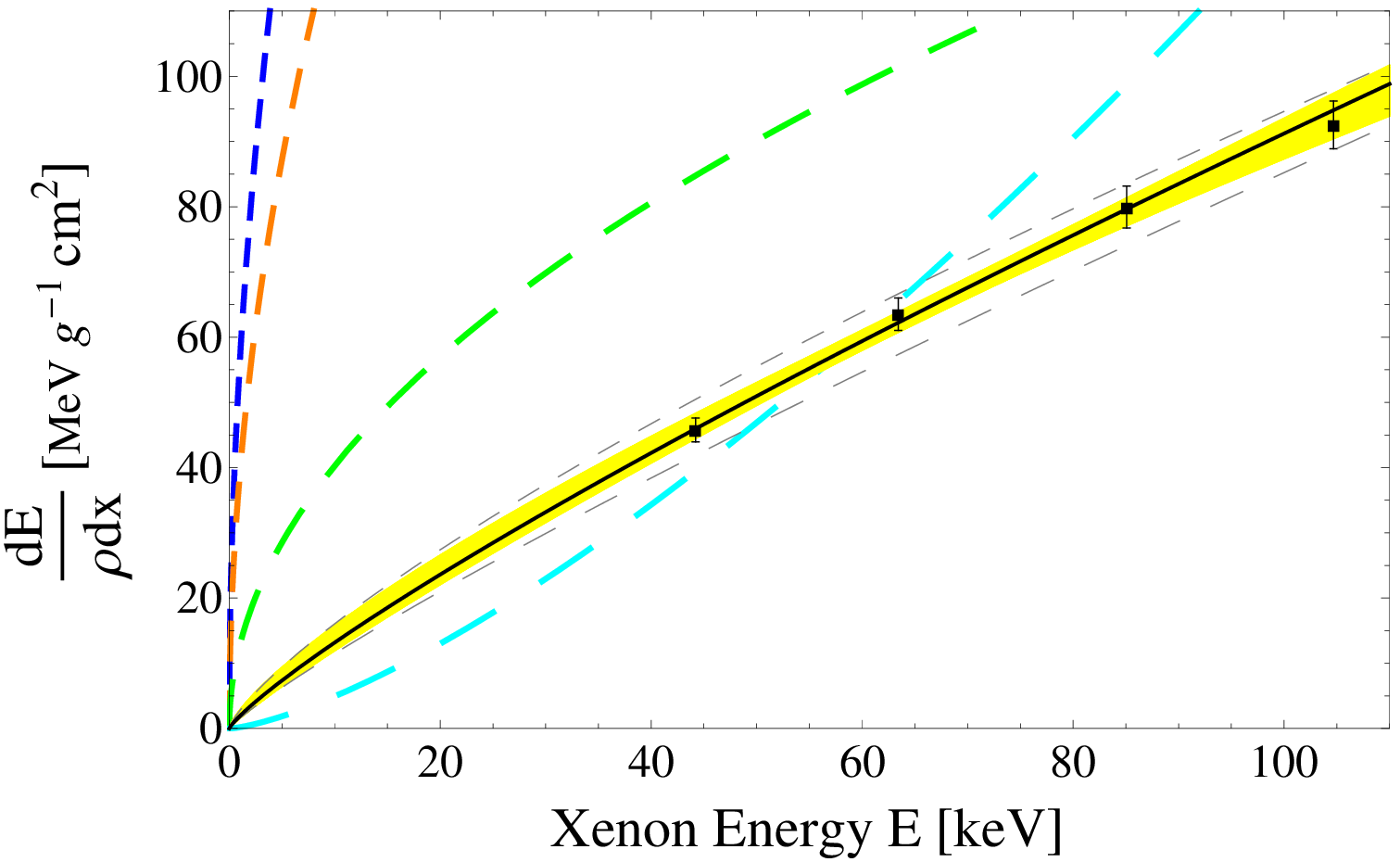}
\includegraphics[width=8cm]{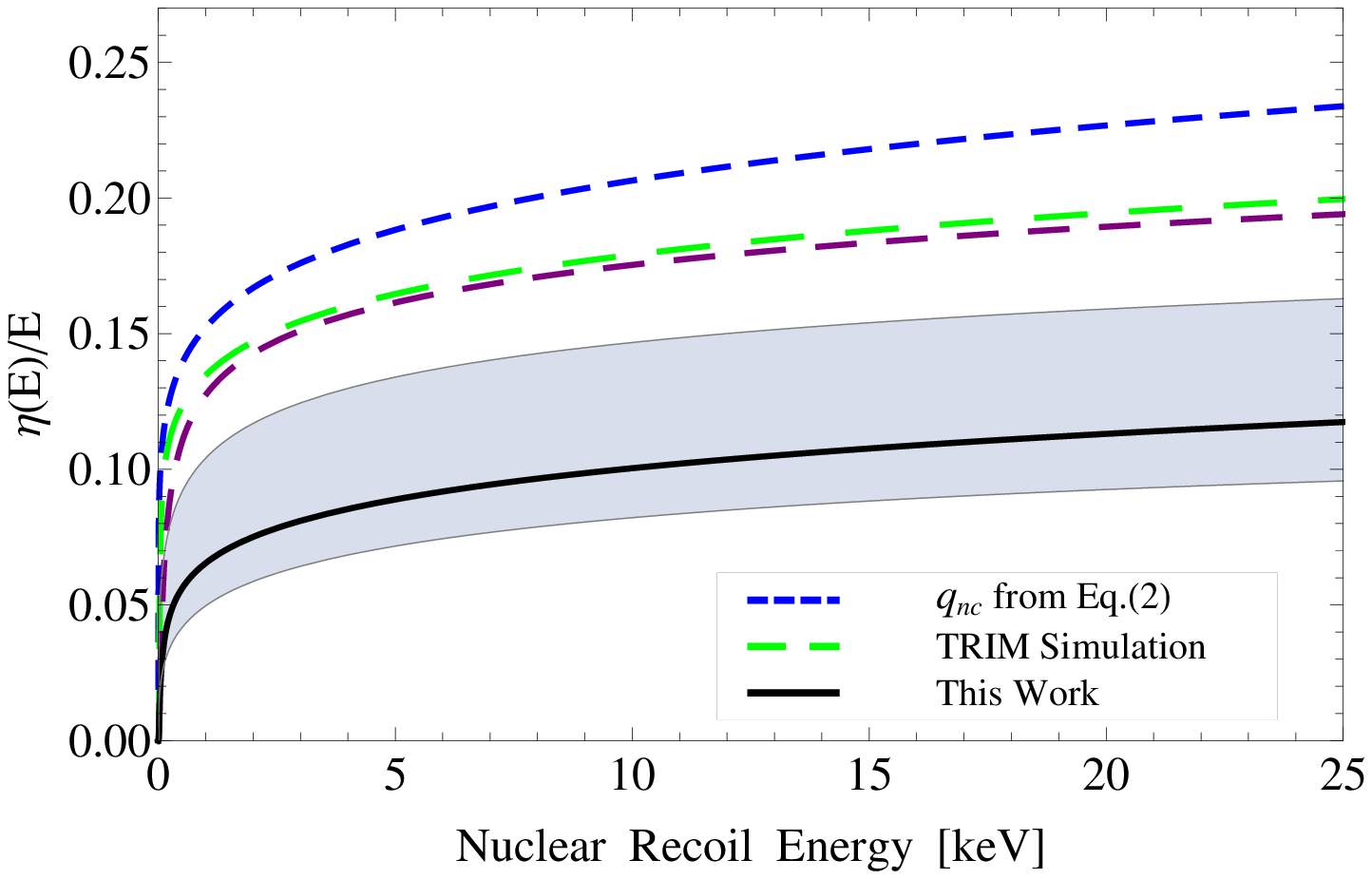}
\caption{Left: Comparisons between different theoretical predictions and experimental data~\cite{Fukuda:1981} on the ESP for a xenon atom in LXe. The ESP fitted from Fukuda's measured data is shown as the solid line and the $2\sigma$ bound around it. Right:
The $q_{\rm nc}$, of LXe obtained from different theoretical predictions. The lower black solid curve and shade is the result from our calculation by using the electronic stopping power fitted to the available experimental measured data.}
\label{fig:SeComp1}
\end{center}
\end{figure}

In principle, ESP as a function of the projectile's kinetic energy can be measured experimentally. However, there is little data in the low-energy region of our interest ($\sim$keV). We have to rely on theoretical consideration and limited data at moderate energies to make progress. Theoretical study of ESP for a charged particle moving in media has been a great interest for over a century~\cite{bohr}. Due to complicated physical processes, theoretical results are scattered. Both the Lindhard-Scharff~(L-S) theory~\cite{Lindhard:1961zz} and Brandt-Kitagawa~(B-K) theory predicted ESP with linear dependence on velocity.  However, Tilinin's study shows that the stopping power decreases much faster, like $v^3$~\cite{tilinin}. Ziegler {\it et al.}~\cite{Ziegler:2010} proposed a semi-empirical result and summarized in a Stopping/Range Table, which is also proportional to velocity at low-energy. As shown in Fig.~{\ref{fig:SeComp1}}, the SRIM data is substantially lower than the predictions of LS and BK theories, but higher than that of Tilinin.

We have also shown the available experimental data in Fig.~{\ref{fig:SeComp1}} (left panel), which was measured in gas phase at low energy region (40$\sim$200~keV)~\cite{Fukuda:1981}. Since the Van der Waals forces are relatively weak for noble gases, the ESP in the liquid phase will be close to that in the gas phase~\cite{Ziegler:2010}. The data indicates a faster fall off at lower energy than $v$, but not as faster as Tilinin's.  Assuming the ESP drops following the same trend in the concerned energy region as that measured at higher energy with a power law behavior~$E^\alpha$, we fit the power-law exponent and the coefficient to the lowest four measured points, and extrapolate the result to the low-energy region. The best fit yields,
\begin{equation}
S_e (E)=1.906 \times E^{0.84} ~{\rm MeV\cdot g^{-1}\cdot cm^2}\ .
\end{equation}
The result is also shown in Fig.~{\ref{fig:SeComp1}} as the black solid line, which lies in between Ziegler's result and Tilinin's. 
As an estimate of the uncertainty, we also generate the 2$\sigma$ error region by randomly generating series ESP functions and using $\chi^2$ distribution to pick out the ones within the 2$\sigma$ error region (yellow band in Fig.~{\ref{fig:SeComp1}}). Then we determine a range of extrapolations defined by two boundary curves,  $S_e^{'} (E)=2.729 \times E^{0.77}$ and $S_e^{''} (E)=1.551 \times E^{0.87}$, which bound the $2\sigma$ region of the data.

Using the above result, we calculate $q_{\rm nc}$ through Monte Carlo simulations. Through 10,000 scattering events, we get a fair average for a discrete set of energies range 0.5$\sim$25~keV, which has a phenomenological fit:
\begin{equation}
q_{\rm nc}(E_{\rm nr})=\frac{e^{-0.033E_{\rm nr}^{-0.958}}}{1+13.789E_{\rm nr}^{-0.189}} \ .
\end{equation}
We plot it in Fig.~1 (right panel) as the black solid line. We have also shown the $2\sigma$ band calculated with the $2\sigma$ fit of the ESP. To calibrate our code, we perform simulations using the SRIM ESP with TRIM parameters in attempt to reproduce the TRIM results. We plot the original TRIM result as the green dashed curve and our reproduction in purple dashed curve. The small difference is due to using of realistic atomic distribution in LXe. We also plot the widely used Lindhard factor~\cite{Mei:2007jn,Hitachi:2005,Bezrukov:2011} as the blue dashed curve for comparison, using the most quoted $k=0.166$. Our $q_{\rm nc}$ is smaller, but not that much smaller, than the quoted Lindhard factor and the TRIM result although the corresponding ESPs are much bigger. This is because the integral equation approach is questionable at low-energy, whereas TRIM uses an artificial cut-off energy beyond which all atom kinetic energy goes to heat. The important feature of our result is that $q_{\rm nc}$ is smaller than ${\cal L}_{\rm eff}$, consistent with the phenomenological observation that the scintillation efficiency for nuclear recoils is larger than that of the electron recoils.

After all recoiling atoms thermalized, the nuclear recoils, similar to electron recoils, produce a certain number of excitons ($N_{ex}$) and electron-ion pairs ($N_i$), where the excitons' relaxation and the electron-ion pairs' recombination will yield scintillation signals. However, some electrons may escape from recombination even under zero electric field, and only a fraction $r$ of electrons are recombined with ions, which exists for both electron recoils and nuclear recoils and must be taken into account when calculating ${\cal L}_{\rm eff}$. In this case, the electronic energy resulting in scintillation, $\eta_{\rm sc} (E_0 )$, is
\begin{equation}
\eta_{\rm sc} (E_0 )=W_{ex} N_{ex}+r\times W_i N_i=W N_i \left(r+\frac{N_{ex}}{N_i} \right) \ ,
\end{equation}
where $W_{ex}$ and $W_i$ are the average energy to produce one exciton and one electron-ion pair, and $W$ is the average energy to produces one scintillation photon, {\it all of which are assumed to be the same for electron recoils and nuclear recoils} as there is no strong evidence otherwise. Once the recombination probability $r$ and the ratio $N_{ex}/N_i$ are determined, ${\cal L}_{\rm eff}$ for a particular electronic energy loss can be calculated.

Thomas and Imel box model was originally proposed to explain electric field dependence of electron-ion recombination in liquid argon or xenon for electron recoils~\cite{Thomas:1987pa}. Through various generalizations, this model successfully describes the electron-ion pair recombination in LXe for nuclear recoils~\cite{Dahl:2009pt,Sorensen:2011,Bezrukov:2011}. In this model, the electron-ion recombination rate depends on the external electric field $\cal E$ as well as the free-charge density $N_0$,
\begin{eqnarray}
r=1-\frac{N_i^{\rm esc}}{N_i} =1-\frac{1}{\xi}\ln(1+\xi)~~~~~\xi=\frac{N_0 \alpha}{4a^2 \mu_- \cal E} \ ,
\end{eqnarray}
where $\alpha/(4a^2 \mu_-)$ is a constant determined by the recombination coefficient $\alpha$, the electron mobility coefficient $\mu_-$, and the ionization-volume length scale $a$. However, even without applied field, there are random fluctuations of electric fields acting on the free electrons, which can also cause electrons to diffuse out of the positive ion region. Extending the model to zero external field, $\xi$ depends on only the ion density $N_0$ in the localized region, which should be linearly proportional to LET at low-energy. Hence we define $\xi\equiv K \times {\rm LET}$, where $K$ is a free parameter obtained phenomenologically from the LET dependence of the scintillation yields in LXe~\cite{Doke:1988nb,Doke:2002}.

The value for $N_{ex}/N_i$ varies from 0.06 to 0.2 for LXe~\cite{Platzman:1961,Takahashi:1975,Doke:2002,Aprile:2010pj},
where 0.06 is the theoretical value~\cite{Takahashi:1975} and 0.2 the measured one~\cite{Doke:2002} for electron recoils. In Ref.~\cite{Aprile:2011prl},
$N_{ex}/N_i=1.07$ or 1.09 is claimed for nuclear recoils from the direct charge measurement. However, these values are fitted to the
data based on much larger $q_{\rm nc}$ in Ref.~\cite{Mei:2007jn,Hitachi:2005,Bezrukov:2011}, which is inconsistent with the
requirement  $q_{nc}\le \cal L_{\rm eff}$. A universal $N_{ex}/N_i$ is also consistent with more scintillation yield from the nuclear recoils
because of the higher recombination rate. When using $N_{ex}/N_i=0.2$, we obtain $K=2.53$~MeV$^{-1} \cdot$g$\cdot$cm$^{-2}$.

\begin{figure}[hbt]
\begin{center}
\includegraphics[width=8cm]{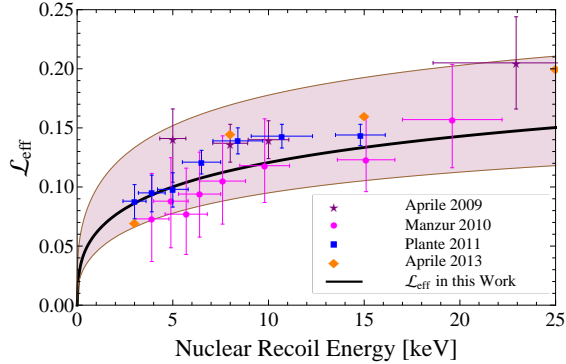}
\caption{The relative scintillation efficiency $\mathcal L_{\rm eff}$ obtained from our calculation, compared with the available experimental data.
The shaded band shows the system and statistical uncertainties with ${\pm}2\sigma$.}
\label{fig:result}
\end{center}
\end{figure}

To get result for ${\cal L}_{\rm eff}$, we convert the electronic energy into scintillation energy taking into account the above 
scintillation quenching. The final result for ${\cal L}_{\rm eff}$ shown as the solid line in Fig.~{\ref{fig:result}} and is 
phenomenologically expressed as:
\begin{eqnarray}
&& {\cal L}_{\rm eff}(E_{\rm nr})=q_{\rm nc}(E_{\rm nr}) \nonumber \\ &&  \times (1.417-0.245\ln(1+4.822E_{\rm nr}^{0.840})E_{\rm nr}^{-0.840})\cdot
\end{eqnarray}

The error band includes both the systematic $2\sigma$ uncertainty from the ESP fit and the statistical error in the simulation. Along with our calculation, we have also shown the experimental data from different measurements. The blue squares are from the Columbia group measurement~\cite{Aprile:2009pc}. The magenta dots are from the Yale group measurement~\cite{Manzur:2010}. The purple stars and the orange diamonds are from the XENON100 collaboration with mono-energy neutron and broad spectrum neutron sources, respectively~\cite{Plante:2011,Aprile:2013teh}. The theoretical result compares very well with the experimental data. The most important message of our calculation, however, is that the scintillation efficiency rapidly decreases with decreasing energy, particularly when less than 3~keV, and goes to zero as the energy goes to zero. This result is expected to be of more general validity because as the recoil energy becomes small, the atom loses its kinetic energy mainly through elastic scattering with other atoms, rather than electronic excitations. This behavior has in some sense already encoded in the ESP.

To conclude, we have made a first realistic study of scintillation efficiency for the low-energy nuclear recoils in LXe. While the study is not completely  {\it ab initio}, it represents the state-of-art theoretical considerations and the result compares favorably with the data from neutron scattering, but differs somewhat from the Monte Carlo fitting to the broad-spectrum neutron data~\cite{Lebedenko:2008gb,Sorensen:2008ec}. The result indicates that LXe scintillation response drops very quickly below 3~keV, a general feature that is independent of many details of the study. We have also used 
the above model including the electric field dependence to study the ionization yield, which can potentially be used either independently or with scintillation together to determine the energy scale of the nuclear recoils. Our preliminary results indicate an excellent agreement with available 
experimental data~\cite{jimuve}.  

{\bf Acknowledgment}
\noindent W. M. thanks Prof. James F. Ziegler for helpful suggestions and discussions. This work is partially supported by a 973 project, No. 2010CB833005, of China's Ministry of Science and Technology, and a grant (No. 11DZ2260700) from the Office of Science and Technology in Shanghai Municipal Government. This work has also been supported by the U.S. Department of Energy via grant DE-AC02-05CH11231,

\end{document}